\documentclass{article}

\usepackage{arxiv}

\usepackage[utf8]{inputenc}  
\usepackage[T1]{fontenc}     
\usepackage{hyperref}       
\usepackage{url}            
\usepackage{booktabs}       
\usepackage{amsfonts}       
\usepackage{nicefrac}       
\usepackage{microtype}      
\usepackage{lipsum}
\usepackage{cite}
\usepackage{graphics}
\usepackage{graphicx}

\title{Insight into Bias in Time-Stratified Case-Crossover Studies}

\author{
  Xiaoming Wang\thanks{Corresponding author. Health Services Statistical and Analytic Methods, Alberta Health Services, \#7235, 2nd Floor, West Wing, Aberhart Centre, 11402 University Avenue, Edmonton, AB T6G 2J3.} \\
  Health Services Statistical and Analytic Methods\\
  Alberta Health Services\\
  Edmonton, AB T6G 2J3 \\
  \texttt{xiaoming.wang@ahs.ca} \\
   \And
 Sukun Wang \\
  Financial Crime Unit\\
  Bank of Montreal\\
  Toronto, ON, Canada \\
  \texttt{sukun492@ualberta.ca} \\
}

\begin{document}
\maketitle

\begin{abstract}
The use of case-crossover designs has become widespread in epidemiological and medical investigations of transient associations. However, the most popular reference-select strategy, the time-stratified schema, is not a suitable solution for controlling bias in case-crossover studies. To prove this, we conducted a time series decomposition for daily ozone (O$_3$) records; scrutinized the ability of the time-stratified schema on controlling the yearly, monthly and weekly time trends; and found it failed on controlling the weekly time trend. Based on this finding, we proposed a new logistic regression approach in which we did adjustment for the weekly time trend. A comparison between the traditional model and the proposed method was done by simulation. An empirical study was conducted to explore potential associations between air pollutants and AMI hospitalizations. In summary, time-stratified schema provide effective control on yearly and monthly time trends but not on weekly time trend. Therefore, the estimation from the traditional logistical regression basically reveals the effect of weekly time trend, instead of the transient effect. In contrast, the proposed logistic regression with adjustment for weekly time trend can effectively eliminate system bias in case-crossover studies.
\end{abstract}


\section{Introduction}
Since the initial work by Malcolm Maclure in 1991 \cite{ref1}, the use of case-crossover designs has become widespread in epidemiological and medical investigations of transient associations between risk factors and adverse health events\cite{ref2,ref3,ref4}, notably in the area of ambient air pollution research\cite{ref5, ref6, ref7, ref8,ref9, ref10, ref11, ref12, ref13}. With a case-crossover design, an investigator samples only cases and compares each individual's exposure during a short time period (hazard period) just before onset of a case event with exposures at other times (referent periods) in a referent window\cite{ref14}.

Comparing with general case-control study designs and cohort study designs, these self-matching designs have an obvious advantage of controlling by design all time-invariant confounders, either measured or unmeasured. The challenge to the case-only design comes from how to control time-varying factors. Being similar to a cohort or a case-control study, a confounding issue in a self-matching design arises when there exists unbalanced matching in determinants between hazard periods and referent periods, leading to various sources of systematic bias. These sources were systematically reviewed by Mittleman and Mostofsky\cite{ref15}.

To control time-varying factors, various reference-select strategies have been proposed in literature. Janes et al.\cite{ref2} provided a comprehensive review of these strategies and assessed potential bias associated with them. These strategies aim to limit the reference-select window to a short period that restricts time-varying confounders to be nearly constant across reference-select windows\cite{ref14}. The ambi-directional\cite{ref16} and the time-stratified\cite{ref17}, were among the top two reference-select strategies in literature. And the latter was claimed to be unbiased\cite{ref2} and optimal\cite{ref18}, avoiding bias resulting from time trend in the exposure series as well as specific time-varying confounders.

As pointed by Wang et al.\cite{ref4} in a case-crossover study, there are two kinds of biases. One is point estimate bias from imperfectly controlling of time-varying factors (confounding issue). The other is the estimation bias of standard error (SE), the so-called overlap bias\cite{ref19,ref20,ref21} from correlations among observations. We could reduce bias in point estimates by using a shorter reference-select window which results in higher overlap bias because shorter reference-select window results in higher correlation among cases and their matched references. In that paper, the authors pointed out the time-stratified case-crossover design is by no means a final solution to the overlap bias. As a remedy, a calibration strategy based on permutation\cite{ref22,ref23,ref24,ref25} was proposed to overcome the overlap bias issue. One pitfall of this strategy is its computational burden. 

In this paper, we were seeking a better solution to the overlap bias in case-crossover studies. For this purpose, we scrutinized the ability of the time-stratified schema on controlling time-varying factors and found its failure on controlling weekly time trend. Based on this finding, we further propose a logistic regression model in which we do adjustment for weekly time trend. As we hypothesized, this model can effectively eliminate the overlap bias as well as potential bias in point estimate. This implies that the overlap bias issue intrinsically roots from unbalanced matching between hazard periods and referent periods of the time-series under investigate. It allows us to have a deep insight into the collection between bias in point estimate and bias in SE estimate.

\section{Insight into Bias in Case-Crossover Studies}
\subsection{Overlap bias}
The term “overlap bias” was first introduced to case-crossover studies by Lumley and Levy\cite{ref17}, who observed that this bias is similar to the “friend control bias” in matched case-control designs\cite{ref20,ref21}.  In the time-stratified design in our simulation study, for example, a hazard day could be viewed as a random sample from a study period, and 3-4 referent days matched are days in the same year, same month and matching weekday. Selecting of referent days is conditioned on hazard days. This reference-select scheme makes the hazard days and the referent days to be selected from different strata in the study period, causing a positive correlation between exposure levels on hazard and referent periods. Austin et al.\cite{ref20} showed that choosing friends or siblings as the matched controls for each case can lead to the so-called overlap bias. 

The deep reason for the overlap bias roots from the correlation among matched cases and controls and leads to SE estimation bias and therefore a biased significance test of the exposure effect. To answer of where the correlations among cases and their matched references come from, we investigated it using an air pollution time-series data.

\subsection{Limitation of time-stratified schema}
We use the same air pollution data used by Wang et al.\cite{ref4}. Briefly, daily records for five criteria air pollutants (in urban area of Edmonton, Canada) spanning a period of 10 fiscal years (from April 1, 2000 to March 31, 2010) were obtained from National Ambient Pollution Surveillance (NAPS) Database\cite{ref26}. The five pollutants were carbon monoxide (CO), nitrogen monoxide (NO), nitrogen dioxide (NO$_2$), ozone (O$_3$), and particulate matter with an aerodynamic diameter $\le 2.5$ (PM$_{2.5}$). Because different pollutants have different units, we standardized each of the five pollutant measures by its interquartile range (IQR, difference between the 75th and the 25th percentile).

\begin{figure*}
\centering
\includegraphics[width=15cm, height=12cm]{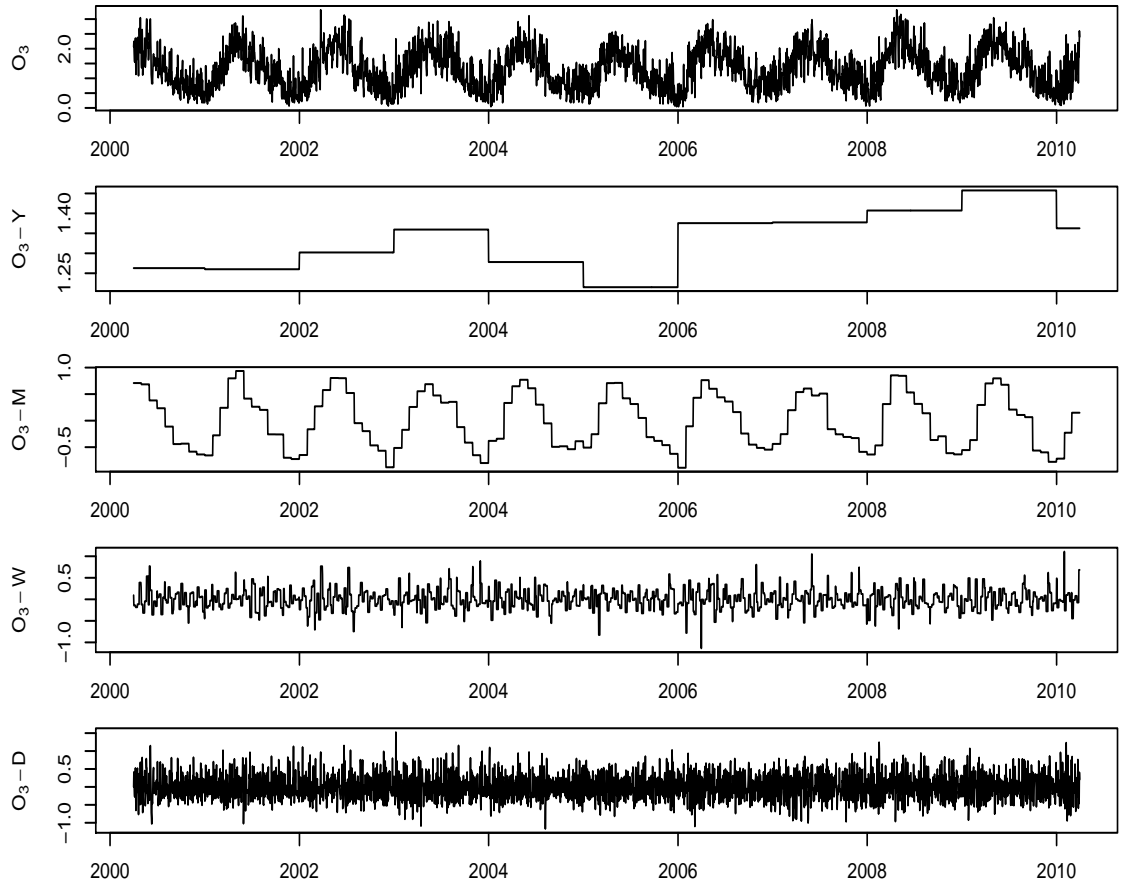}
\caption{Time series decomposition of daily records of ozone (O$_3$) from April 1, 2000 to March 31, 2010. Daily records of O$_3$ (in Edmonton urban area, Canada) are depicted in the top panel. Its yearly, monthly, weekly, and daily components of time trends are depicted in 2nd to 5th panels, respectively.}
\label{fig1}
\end{figure*}

To see the limitation of the time-stratified reference-select schema, we conducted a time-series decomposition for the daily records of O$_3$. From the daily time-series of O$_3$ we calculated its yearly average (O$_3^y$), monthly average after removing the yearly time trend (O$_3^m$), weekly average after removing the yearly and monthly time trends (O$_3^w$) and daily average after removing the yearly, monthly and weekly time trends (O$_3^d$). The above process is an usual time-series decomposition, such that $O_3=O_3^y+O_3^m+O_3^w+O_3^d$. We have depicted the time-series and its components in Figure \ref{fig1}. 

\begin{center}
(Figure 1 here)
\end{center}

From Figure \ref{fig1}, we see that (a) the time-stratified schema can control the yearly and monthly time trends of O$_3$, because a case and its matched references (with 3 to 4 references) are always in the same year and the same month, in which yearly and monthly average values of O$_3$ are constant; and (b) the time-stratified schema can do nothing on controlling weekly time trend of O$_3$.

The transient effect of an air pollutant on a health outcome is a within-week impact of the air pollutant. We believe that a lack of controlling weekly time trend is the main source of potential point estimation bias as well as overlap bias. Correlations among a case and its matched references could not be ignorable if the weekly trend time-series is highly correlated.

\section{Correct Bias in Case-Crossover Studies}
Suppose $Y$ is the response variable where $Y=0/1$ indicates exposure to hazard/referent period in a case-crossover study. The following traditional logistic regression model is the typical analytic tool being used in literature to assess transient association between adverse events and exposure to a pollutant. 
\begin{equation}
\mbox{logit}(p)=  \alpha + \beta O_3,
\end{equation}
here $\mbox{logit(p)}=\log(p/(1-p)$ with $p=Pr(Y=1)$, and $\beta$ is the transient effect of O$_3$ under estimation. Again, we use O$_3$ as example for an interested pollutant. For illustration simplicity, we consider a model without covariates.

\subsection{Control weekly time trend}
We propose the following analytic methods. Given a suitable model adjustment for weekly, monthly and yearly time trends, we seek to obtain unbiased point estimate and unbiased significance test. Our first logistic regression model is the one adjusted for the yearly, monthly and weekly time trends of O$_3$, i.e., 
\begin{equation}
\mbox{logit}(p)=  \alpha + \beta O_3 +\gamma_1 O_3^y+\gamma_2 O_3^m +\gamma_3 O_3^w. 
\end{equation}
The adjustment for the yearly and monthly time trends is mainly because some cases matched with 3 references while the others matched with 4 references, leading to possible unbalanced matching of the yearly and monthly time trends in hazard and reference periods.

The second logistic regression model is designed to estimate directly the impact of the daily average (O$_3^d$) and adjusted for the yearly, monthly and weakly time trends.
\begin{equation}
\mbox{logit}(p)=  \alpha + \beta O_3^d +\gamma_1 O_3^y+\gamma_2 O_3^m +\gamma_3 O_3^w.    
\end{equation}
The transient impact from O$_3$ is considered to be the effect of O$_3^d$. It is worth to point out that model 2 and 3 are equivalent with each other. 

If model 2 and 3 are correct, i.e., there is no other unobserved time-varying factors confounding the association under estimation, we can conclude that both of them can provide unbiased point estimate (regardless of whether the estimate of SE is biased or not) and unbiased significance test (if estimate of SE is unbiased). 

Generally speaking, if we cannot control the weekly time trends, correlations among weekly average records will enter into the errors, leading to relatively higher correlations among them. However, after controlling weekly time trend, correlations among daily average records could be low for days in the same month but with one week or multiple weeks apart. If we can assume that the daily average records in a hazard day and its matched reference days are independent with each other, then we can conclude the significance test from model 2 or 3 is unbiased.

\subsection{Calibrate bias using permutation}
From statistical point of view, all of the proposed models are only for ``univariate analysis". However, bias could also come from other potential time-varying confounders\cite{ref13}. An idea strategy is to observer those time-varying confounders and do suitable model adjustment for them. In this way, model (2) can be rewritten as
\begin{equation}
\mbox{logit}(p)=  \alpha + \beta O_3 +\gamma_1 O_3^y+\gamma_2 O_3^m +\gamma_3 O_3^w +\mu Z,
\end{equation}
here the item of $Z$ represents suitable adjustment for some observed covariates. However, a big challenge is it is impossible for an investigator to observer all the potential time-varying confounders.

Another possible (but not perfect) solution to the challenge is using the calibration method suggested by Wang et al.\cite{ref4} (see detail of the calibration algorithm therein). The basic assumption of the calibration method is that the estimated exposure effect ($\hat{\beta}$) can be always divided into design bias($b$), true exposure effect ($\beta$), and random error ($e$), i.e. 
\begin{equation}
\hat{\beta}=\beta+b+e,                 
\end{equation}
which means regardless of whether the true exposure effect $\beta$ is zero or not, the design bias $b$ will always be a constant. 

Under this assumption, we can correctly estimate the design bias $b$ by permutation and completely eliminate it from point estimate $\hat{\beta}$. Although this assumption may be questionable in practice, we still believe this non-parametric technique can gain benefit for us on bias reduction in both point estimate and significance test. We will show performance of our calibration strategy by simulation. 

\section{Simulation Study}
We use O$_3$ as an example to describe our simulation study design. The same design applies to all the five pollutants. Based on the 3,652-day study period (from April 1, 2000 to March 31, 2010), we simulated “adverse event days” to create different scenarios to show the performance of the following modeling strategies. 
\begin{itemize}
   	\item Traditional logistic model 1;
	\item Proposed logistic model 2;
	\item Traditional model 1 with calibration;
	\item Proposed model 2 with calibration.
\end{itemize}

For each scenario, we replicate 1,000 experiments. In each of these experiments: 5,000 event days are randomly sampled with replacement from the study period with probability proportional to $\exp(\beta O_3^d+\gamma_1 O_3^y+\gamma_2 O_3^m+\gamma_3 O_3^w)$; references are selected using the time-stratified schema; data are analyzed using the above four modeling strategies. All analyses were conducted in R (version 3.6.1, 2019-07-05).

\subsection{Bias check and size check under null}
\begin{figure*}
\centering
\includegraphics[width=15cm, height=12cm]{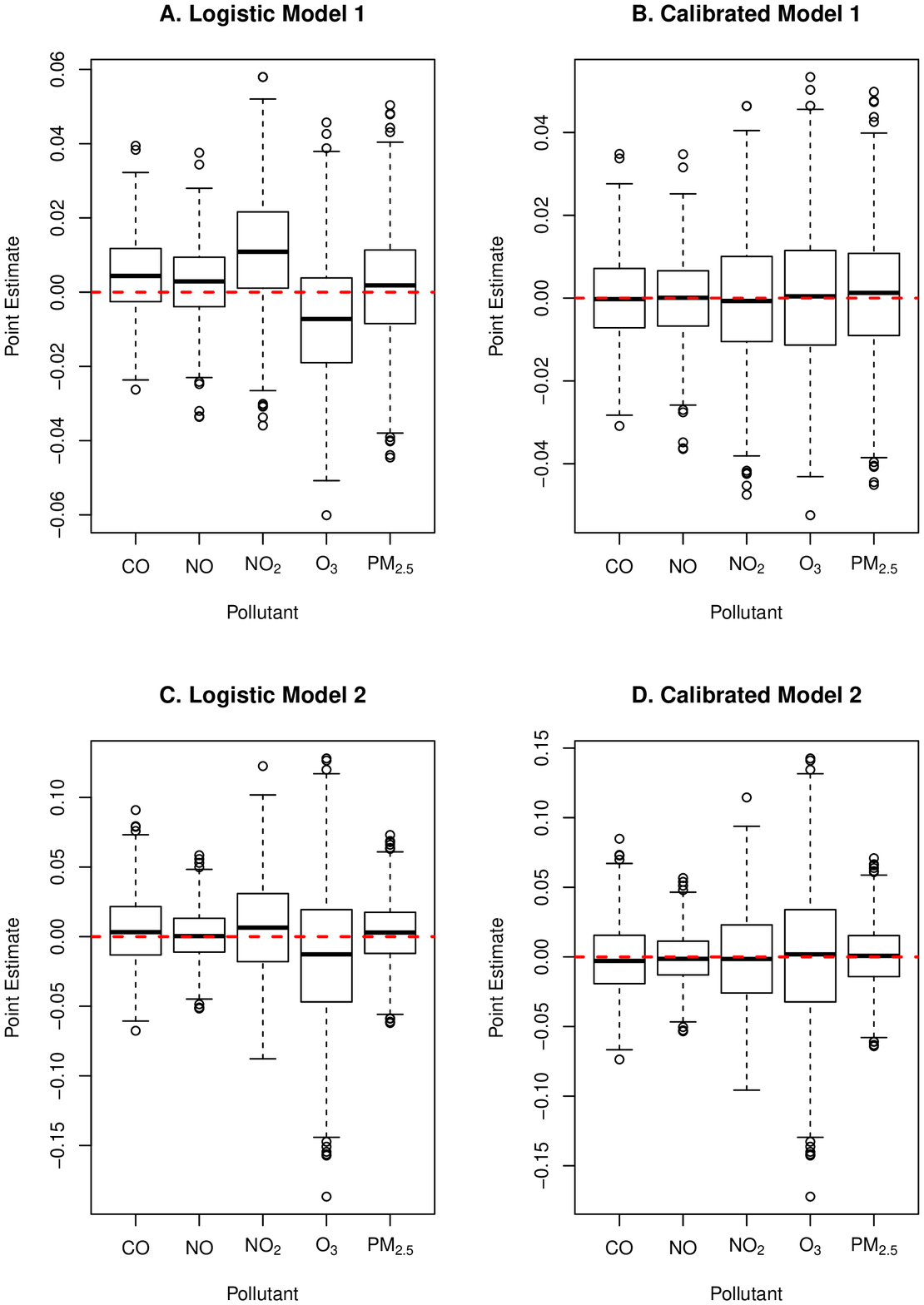}
\caption{Boxplot of distribution the estimated biases of effect of one of the five pollutants using one of the four analytic modeling strategies respectively. Scenarios are simulated under $\beta=0$ and $\gamma=0$, and study period is the whole-cycle-period from April 1, 2000 to March 31, 2010. The five pollutants are carbon monoxide (CO), nitrogen monoxide (NO), nitrogen dioxide (NO$_2$), ozone (O$_3$), and particulate matter with an aerodynamic diameter $\le 2.5$ (PM$_{2.5}$).  }
\label{fig2}
\end{figure*}

\begin{figure*}
\centering
\includegraphics[width=15cm, height=12cm]{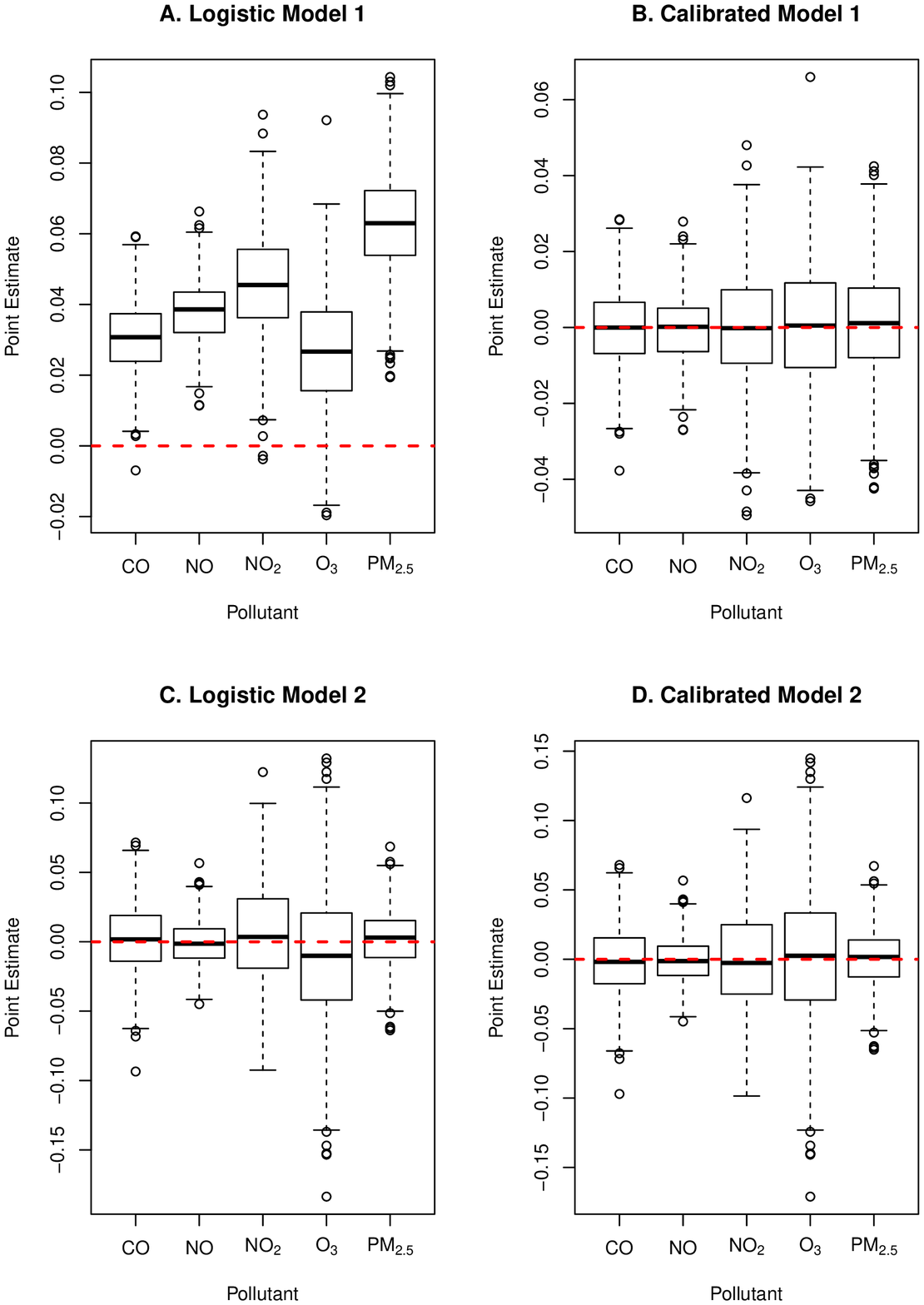}
\caption{Boxplot of distribution of the estimated biases of effect of one of the five pollutants using one of the four analytic modeling strategies respectively. Scenarios are simulated under $\beta=0$ and $\gamma=0.2$, and study period is the whole-cycle-period from April 1, 2000 to March 31, 2010. The five pollutants are carbon monoxide (CO), nitrogen monoxide (NO), nitrogen dioxide (NO$_2$), ozone (O$_3$), and particulate matter with an aerodynamic diameter $\le 2.5$ (PM$_{2.5}$).  }
\label{fig3}
\end{figure*}

All scenarios in this subsection are designed under null hypothesis ($\beta=0$) with the settings of $\gamma_1=\gamma_2=\gamma_3=\gamma=0$ and 0.2, respectively. For each of the five pollutants, point estimates from each of the four analytic methods are depicted in Figure \ref{fig2} and Figure \ref{fig3}, and rejection frequencies (size, under the criterion $\alpha_0=0.05$) of significance tests in 1,000 experiments are reported in Table \ref{tab1}.

\begin{center}
(Table 1 here)
\end{center}

From the top panel of Table \ref{tab1}, we can see that in scenarios under the setting of $\gamma = 0$, the size of model 1 (before calibration) is far less from the nominal level 50, especially for CO, NO, and O3. Considering the point estimates in these situations are almost unbiased (Figure \ref{fig2}-A), we conclude that the bias in size is from correlation among errors (i.e., overlap bias). This implies that, without controlling weekly time trend, model 1 provides overestimate of standard errors, leading to higher overlap bias. On the other hand, with suitable adjustment on weekly time trend (model 2), the overlap bias is almost gone and significance tests nearly unbiased. It seems that the calibration method can always provide reasonable and robust significance tests. 

The bottom panel of Table \ref{tab1} shows that under the setting of $\gamma = 0.2$ there is a big problem in using model 1 (before calibration). The big size of significance tests is from the uncontrolled weekly time trend, which leads to seriously biased point estimates (Figure \ref{fig3}-A) as well as a higher correlation among errors. To our surprise, in this scenario, the calibration strategy also can do a great job for overlap bias correction. 

The proposed logistic regression model 2 is unbiased in significance test both before and after calibration. The good performance of model 2 roots from thoroughly controlling of the weekly time trend. In this case, the calibration method can do very little on further bias reduction for model 2.

\begin{center}
(Figure 2 here)
\end{center}

Figure \ref{fig2} shows the performance in point estimates of the traditional logistic model 1 vs the proposed logistic model 2 under the setting of $\gamma=0$. Because of the null hypothesis, a point estimate is its estimation bias. In this scenario point estimates from model 1 are nearly unbiased (panel A); after calibration (panel B) bias in point estimates is eliminated completely. It is obvious that model 2 provides very similar results on estimation bias before and after calibration (panel C and D), i.e., unbiased point estimates. 

Comparing the top panels (A and B) and the bottom panels (C and D) in Figure \ref{fig2}, we can see a difference in variance of estimation bias: range of top panels is within $\pm 0.06$, while that of bottom panels is within $\pm 0.15$. This can be explained by the difference between model 1 and 2. In model 1, the time-stratified schema is used to control yearly and monthly time trends, so the estimated effect is from weekly time trends. In contract, with controlling yearly, monthly and weekly time trends, the estimated effect from model 2 is that of the daily time trend. In fact, the target estimation of model 1 is the effect of weekly time trend, instead of the transient effect under investigation.

\begin{center}
(Figure 3 here)
\end{center}

Under the setting of $\gamma=0.2$, model 1 and model 2 perform quite different in point estimates (Figure \ref{fig3}). It is obvious that (a) without control of weekly time trend, model 1 could be unacceptably biased in coefficient estimate (panel A); (b) with thoroughly control of weekly time trend, model 2 can be perfectly unbiased in point estimates (panel C); (c) the calibration method can eliminate the bias from model 1 by permutation (panel B); and (d) model 2 and calibrated model 2 perform very similar, reflecting that model 2 is good enough in point estimate and further calibration is unnecessary in the simulated scenario.

\subsection{Power check under alternative}
In order to compare efficiency, we considered a power check for the above four analytic strategies. Scenarios are simulated by setting $\gamma=0$ and $\beta$ from ${0,0.1,\cdots,0.4}$. Rejection rates (power, under the criterion $\alpha_0=0.05$) of significance tests in 1,000 experiments under each scenario are reported in Table \ref{tab2}.

\begin{center}
(Table 2 here)
\end{center}

In summary, (a) test power form model 1 is the worst, because of the overlap bias; (b) model 2 provides competitive best test power with model 2 with calibration; and (c) model 1 with calibration is much better than model 1 in test power, but worse than model 2 and model 2 with calibration. The difference of the two calibrated models in performance can be explained by the different in their estimation targets (see previous subsection).

\section{Empirical study}

Using Alberta Health administrative databases, we obtained all historical records of hospital admissions for acute myocardial infarction (AMI, ICD-10 code I21-I22 or ICD-9 code 410) for Edmonton urban dwellers. In total 9811 admissions for AMI occurred from 1 April 2000 to 31 March 2010 among patients aged 20 or over and living in the areas of Edmonton.

We explore potential transient associations between AMI hospitalizations and the five air pollutants (the same data used in the simulation study). Cohort and 10 sub-cohorts are defined using admission characteristics as Whole (whole cohort), Male, Female, Age$\ge 65$, Age$<65$, STEMI (ST Segment Elevation Myocardial Infarction), NSTEMI (Non-ST Segment Elevation Myocardial Infarction), Diabetes, Hypertension, Dysrhythmia, and PIHD (prehistory of heart disease); and records in each of them are further divided into subgroups defined by seasons (All season, Spring, Summer, Autumn and Winter). For each subgroup: hazard days are selected as $k$th-day pre case events ($k=0-4$); references are selected using the time-stratified schema; data are analyzed using the traditional logistic regression model 1 as well as the proposed logistic regression model 2.

The estimated effects of the 1,375 (11 cohort/sub-cohorts x 5 subgroups x 5 lags x 5 pollutants) candidate associations from the traditional logistic model 1 and the proposed logistic model 2 are reported in appendix files "Est-AMI-application-model1.csv" and "Est-AMI-application-model2.csv", respectively.

From the estimates of model 1, in summary, 793 (582) of them were estimated as positive (negative) effects; there are 140 (39) candidates with p-value $\le 0.05$ ($0.01$). From the estimates of model 2, on the other hand, 634 (741) of them were estimated as positive (negative) effects; there are 75 (10) candidates with p-value $\le 0.05$ ($0.01$). Using Bonferroni correction with criterion $\alpha_0=0.05/1375$, none of the associations (neither from model 1 nor from model 2) has a p-value less than the criterion. The correlation is 0.605 between estimates of model 1 and estimates of model 2. From these numbers we can see an obvious difference in results between the two models.

\begin{center}
(Table 3 here)
\end{center}

The 39 associations detected by the model 1 with p-value $\le 0.01$ are reported in Table \ref{tab3}. From the table we can see a clear seasonal pattern: CO, NO, and NO$_2$ have positive effects in spring and negative effects in autumn; O$_3$ has a negative effect in winter or Spring, and positive effect in summer. It is hard to believe that the same pollutant in one season has a protective effect while in another season has a harmful effect. This phenomenon implies that estimate of the traditional logistic model 1 contains some kind of seasonal trend. We believe it is just because model 1 can not control the weekly time trend of an exposure time series. On the other hand, model 2 with adjusting with weekly time trend can be expected to tell a different story.

\begin{center}
(Table 4 here)
\end{center}

The 10 associations detected by the model 2 with p-value $\le 0.01$ are reported in Table \ref{tab4}. From the table we see there is no seasonal pattern as shown in Table \ref{tab3}.  In detail, all of the 9 estimated effects of CO, NO, O$_3$ or PM$_{2.5}$ are positive, and 1 estimated effect of NO$_2$ is negative. This implies that exposure to a higher concentration of CO, NO, O$_3$ or  PM$_{2.5}$ may associated with higher risk of AMI hospitalization. For example, STEMI patients could be sensitive to increased PM$_{2.5}$ concentration in autumn (OR=1.156; 95\% confident interval(CI): 1.047, 1.276); while Dysrhythmia patients exposed to increased concentration of PM$_{2.5}$ in spring could have elevated risk of hospitalization (OR=1.247; CI: 1.058, 1.468).

It is notable that there are 4 candidate associations with p-value $\le 0.01$ that were detected by both the two methods. These estimates are marked with an asterisk in Table \ref{tab3} and Table \ref{tab4}, respectively. For example, a higher O$_3$ level in summer is associated with AMI hospitalizations of all patients (OR=1.216; CI=1.061, 1.392), NSTEMI patients (OR=1.308; CI=1.806, 1.576) and hypertension patients (OR=1.281; CI=1.072, 1.531) by the proposed model 2.

\section{Conclusions}

Using the decomposition of the ozone time series, we scrutinized the control ability of the time-stratified reference selecting schema. This schema provides effective control on yearly and monthly time trends but not on weekly trend. We believe that a lack of control on weekly trend is the main source of systematic bias in case-crossover studies. 

Without controlling weekly time trend, it will enter into the errors of our logistic regression model, leading to the so-called overlap bias and a possible bias in point estimate. Doing model adjustment for weekly time trend can be an effective way of reducing bias from it. Further model adjustment and calibration could be necessary if there are other potential time-varying confounders. 

The typical case-crossover study design, i.e., combining the time-stratified reference-select schema with a traditional logistic regression, leads to an estimate of weekly time trend effect, instead of the transient effect under investigation. The difference in results between the traditional approach and the proposed method could be significant. We strongly recommend researchers to do controlling (adjustment) for weekly time trend of exposures and covariates in their case-crossover application studies.

We found, through the empirical study for AMI hospitalization, that higher concentration of CO, NO, O$_3$ and PM$_{2.5}$ are associated with higher risk of AMI hospitalization, while higher concentration of NO$_2$ is associated with lower risk of AMI hospitalization. A limitation of the empirical study for AMI hospitalization is that we only did `univatiate' analyses for the five pollutants without controlling for impact from other potential time-varying factors. Considering also none of the candidate association has p value less than the Bonferroni criterion, we believe further confirmation for the reported ``findings" is still needed from other studies.

\section*{Acknowledgements}
Ethical approval for the study was granted by the University of Alberta’s Health Research Ethics Board-Health Panel (IREB Pro00010852). Patient records/information was anonymized and de-identified with a unique scrambled ID and released to us in this form by the ministry of health prior to analysis. 

\subsection*{Funding}
The author(s) received no financial support for the research, authorship, and/or publication of this article.

\subsection*{Conflicts of Interest}
The author(s) declared no potential conflicts of interest with respect to the research, authorship, and/or publication of this article.

\subsection*{Supplementary Material}
Air pollution data of daily average records in urban Edmonton from April 1, 2000 to March 31, 2010 for carbon monoxide (CO), nitrogen monoxide (NO), nitrogen dioxide (NO$_2$), ozone (O$_3$), and particulate matter with an aerodynamic diameter $\le 2.5$ (PM$_{2.5}$) is available online. An Excel file reporting modeling results of the 1375 models for the empirical study is available online. R code files for the simulation and the empirical study are available upon request.

\newpage
\begin{table}
 \caption{Size of significance tests for model 1-2 (before calibration and after calibrations).}
  \centering
  \begin{tabular}{lcl|ccccc}
\noalign{\smallskip}\hline\noalign{\smallskip}
    Model  & $\gamma$ &Calibration&CO &NO &NO$_2$&O$_3$ &PM$_{2.5}$\\
\hline
    Model 1& 0.0      &Before     &4  &9  &16 &4  &42\\
    Model 1& 0.0      &After      &61 &50 &52 &59 &63\\
    Model 2& 0.0      &Before     &48 &35 &46 &59 &57\\
    Model 2& 0.0      &After      &49 &51 &60 &46 &59\\
 \hline
    Model 1& 0.2      &Before     &617&973&619&108&990\\
    Model 1& 0.2      &After      &54 &63 &58 &54 &51\\
    Model 2& 0.2      &Before     &47 &38 &43 &58 &51\\
    Model 2& 0.2      &After      &54 &55 &59 &63 &63\\
\hline
  \end{tabular}
  \label{tab1}
\end{table}
\noindent\small{Note: Model 1=traditional logistic regression model on pollutant without adjustment; Model 2=proposed logistic regression model on pollutant with adjustment on its yearly, monthly and weekly time trend; each size (under the criterion $\alpha_0=0.05$) was calculated from 1000 independent replication of experiments.}

\vskip 2cm
\newpage
\begin{table}
 \caption{Power of significance tests for model 1-2 (before and after calibrations) with $\gamma= 0$.}
  \centering
  \begin{tabular}{lc|cc|cc}
\hline
     \multicolumn{2}{c}{Scenario} &
     \multicolumn{2}{c}{Before Calibration} &
     \multicolumn{2}{c}{After Calibration} \\
\hline  
    pollutant &$\beta$&Model 1&Model 2&Model 1&Model 2\\
\hline
    CO        &0.0    &4   &48  &61  &49\\
    CO        &0.1    &637 &964 &819 &937\\
    CO        &0.2    &1000&1000&1000&1000\\
    CO        &0.3    &1000&1000&1000&1000\\
 \hline
    NO        &0.0    &9   &35  &50  &51\\
    NO        &0.1    &979 &999 &986 &999\\
    NO        &0.2    &1000&1000&1000&1000\\
    NO        &0.3    &1000&1000&1000&1000\\
 \hline
    NO$_2$    &0.0    &16  &46  &52  &60\\
    NO$_2$    &0.1    &420 &689 &515 &663\\
    NO$_2$    &0.2    &960 &1000&978 &1000\\
    NO$_2$    &0.3    &1000&1000&1000&1000\\
\hline   
    O$_3$     &0.0    &4   &59  &59  &46\\ 
    O$_3$     &0.1    &45  &447 &376 &557\\
    O$_3$     &0.2    &454 &976 &898 &988\\
    O$_3$     &0.3    &930 &1000&1000&1000\\
\hline 
    PM$_{2.5}$&0.0    &42  &57  &63  &59\\
    PM$_{2.5}$&0.1    &980 &999 &979 &998\\
    PM$_{2.5}$&0.2    &1000&1000&1000&1000\\
    PM$_{2.5}$&0.3    &1000&1000&1000&1000\\
\hline
  \end{tabular}
  \label{tab2}
\end{table}
\noindent\small{Note: Model 1=traditional logistic regression model on pollutant without adjustment; Model 2=proposed logistic regression model on pollutant with adjustment on its yearly, monthly and weekly time trend; each power (under the criterion $\alpha_0=0.05$) was calculated from 1000 independent replication of experiments.}
\vskip 2cm

\newpage
\begin{table}
\caption{Detected associations with p-value $\le$ 0.01 between pollutants and AMI hospitalization by the traditional logistic model 1.}
\centering
\begin{tabular}{llccc|c}
\hline
Sub-cohort&Season&N&Pollutant&Lag (day)& Odds Ratio (95\% CI)\\
\hline
Whole & Spring & 2505 & CO & 4 & 1.068 (1.017, 1.121) \\
Age$<$65 & Spring & 1045 & CO & 4 & 1.121 (1.042, 1.205) \\
NSTEMI & Spring & 1397 & CO & 4 & 1.093 (1.022, 1.168) \\
Whole & Autumn & 2408 & CO & 3 & 0.919 (0.865, 0.976) \\
Age$\ge$65 & Autumn & 1320 & CO & 3 & 0.893 (0.822, 0.971) \\
\hline
Whole & Spring & 2505 & NO & 2 & 1.067 (1.021, 1.115) \\
Whole & Spring & 2505 & NO & 3 & 1.066 (1.020, 1.114) \\
Whole & Spring & 2505 & NO & 4 & 1.068 (1.019, 1.119) \\
Male & Spring & 1631 & NO & 3 & 1.081 (1.024, 1.141) \\
Male & Spring & 1631 & NO & 4 & 1.078 (1.019, 1.14) \\
Age$<$65 & Spring & 1045 & NO & 3 & 1.106 (1.034, 1.184) \\
Age$<$65 & Spring & 1045 & NO & 4 & 1.110 (1.034, 1.191) \\
STEMI & Spring & 1108 & NO & 3 & 1.091 (1.024, 1.163) \\
Dysrhythmia* & Spring & 428 & NO & 3 & 1.158 (1.046, 1.282) \\
NSTEMI & Spring & 1397 & NO & 4 & 1.091 (1.024, 1.162) \\
Whole & Autumn & 2408 & NO & 3 & 0.915 (0.861, 0.973) \\
Whole & Autumn & 2408 & NO & 4 & 0.907 (0.852, 0.966) \\
Male & Autumn & 1607 & NO & 2 & 0.906 (0.841, 0.976) \\
Male & Autumn & 1607 & NO & 4 & 0.899 (0.832, 0.972) \\
Age$\ge$65 & Autumn & 1320 & NO & 3 & 0.880 (0.806, 0.96) \\
\hline
Whole & Spring & 2505 & NO$_2$ & 2 & 1.093 (1.030, 1.160) \\
STEMI & Spring & 1108 & NO$_2$ & 2 & 1.138 (1.04, 1.244) \\
Whole & Autumn & 2408 & NO$_2$ & 3 & 0.837 (0.761, 0.921) \\
Whole & Autumn & 2408 & NO$_2$ & 4 & 0.873 (0.794, 0.960) \\
Male & Autumn & 1607 & NO$_2$ & 2 & 0.857 (0.764, 0.96) \\
Male & Autumn & 1607 & NO$_2$ & 3 & 0.805 (0.716, 0.905) \\
Male & Autumn & 1607 & NO$_2$ & 4 & 0.840 (0.748, 0.945) \\
Age$\ge$65 & Autumn & 1320 & NO$_2$ & 3 & 0.828 (0.727, 0.944) \\
HTN & Autumn & 1423 & NO$_2$ & 3 & 0.841 (0.743, 0.953) \\
NSTEMI & Autumn & 1312 & NO$_2$ & 3 & 0.800 (0.703, 0.911) \\
NSTEMI & Autumn & 1312 & NO$_2$ & 4 & 0.830 (0.729, 0.946) \\
\hline
Whole & Winter & 2475 & O$_3$ & 1 & 0.870 (0.783, 0.967) \\
Whole & Spring & 2505 & O$_3$ & 4 & 0.882 (0.815, 0.955) \\
Female & Spring & 874 & O$_3$ & 4 & 0.832 (0.726, 0.954) \\
NSTEMI & Spring & 1397 & O$_3$ & 4 & 0.870 (0.783, 0.966) \\
Whole* & Summer & 2423 & O$_3$ & 2 & 1.144 (1.037, 1.261) \\
NSTEMI* & Summer & 1288 & O$_3$ & 2 & 1.204 (1.052, 1.377) \\
HTN* & Summer & 1433 & O$_3$ & 2 & 1.183 (1.041, 1.345) \\
\hline
Age$<$65 & Spring & 1045 & PM$_{2.5}$ & 4 & 1.107 (1.028, 1.191) \\
\hline
\end{tabular}
\label{tab3}
\end{table}
\noindent\small{Note: N=number of observations; CI=confidence interval; STEMI = ST Segment Elevation Myocardial Infarction; NSTEMI = Non-ST Segment Elevation Myocardial Infarction; HTN = Hypertension; PIHD = prehistory of heart disease; *Detected by both the traditional logistic model 1 and the proposed logistic model 2.}

\newpage
\begin{table}
\caption{Detected associations with p-value $\le$ 0.01 between pollutants and AMI hospitalization by the proposed logistic model 2.}
\centering
\begin{tabular}{llccc|c}
\noalign{\smallskip}\hline\noalign{\smallskip}
Sub-cohort&Season&N&Pollutant&Lag (day)& Odds Ratio (95\% CI)\\
\hline
Dysrhythmia & Spring & 428 & CO & 3 & 1.319 (1.094, 1.591)\\
Dysrhythmia* & Spring & 428 & NO & 3 & 1.210 (1.048, 1.397)\\
HTN & All Season & 5848 & NO$_2$ & 0 & 0.912 (0.850, 0.978)\\
Whole* & Summer & 2423 & O$_3$ & 2 & 1.216 (1.061, 1.393)\\
NSTEMI* & Summer & 1288 & O$_3$ & 2 & 1.308 (1.086, 1.576)\\
Diabetes & All Season & 2667 & O$_3$ & 3 & 1.192 (1.047,1.356)\\
HTN* & Summer & 1433 & O$_3$ & 2 & 1.281 (1.072, 1.531)\\
PIHD & All Season & 3076 & O$_3$ & 3 & 1.174 (1.040, 1.325)\\
STEMI & Autumn & 1096 & PM$_{2.5}$ & 4 & 1.156 (1.047, 1.276)\\
Dysrhythmia & Spring & 428 & PM$_{2.5}$ & 3 & 1.247 (1.058, 1.468)\\
\hline
\end{tabular}
\label{tab4}
\end{table}
\noindent\small{Note: N=number of observations; CI=confidence interval; STEMI = ST Segment Elevation Myocardial Infarction; NSTEMI = Non-ST Segment Elevation Myocardial Infarction; HTN = Hypertension; PIHD = prehistory of heart disease; SE=standard error; *Detected by both the traditional logistic model 1 and the proposed logistic model 2.}

\newpage

\newpage

\newpage

\end{document}